# Long-range forces extending from polymer-gel surfaces

Jian-ming Zheng and Gerald H. Pollack

Dept. of Bioengineering, Box 357962

University of Washington

Seattle WA 98195

Contact Information   email:  *ghp@u.washington.edu*
                      FAX:  (206) 685-3300

# Abstract

Aqueous suspensions of microspheres were infused around gels of varying composition. The solutes were excluded from zones on the order of 100 μm from the gel surface. We present evidence that this finding is not an artifact, and that solute-repulsion forces exist at distances far greater than conventional theory predicts. The observations imply that solutes may interact over an unexpectedly long range.

## Introduction

Solute-solute interactions in aqueous solution are generally thought to occur on a scale of nanometers. At such close range, molecules exhibit either repulsive or attractive interactions depending on the nature of the solutes and solvent, and on the magnitude of separation (Israelachvili, 1992; Parsegian, 2002). Beyond those distances, forces between solutes are expected to vanish, especially when the solution contains salts to mask any surface charges that may be present.

On the other hand, several reports imply that solutes are influenced by the presence of hydrophilic surfaces at relatively macroscopic distances (Kepler and Fraden, 1994; Crocker and Grier, 1996; Xu and Yueng, 1998). In those reports, the behavior of solutes in confined spaces, or, at distances up to 1 μm or more from a surface, differs from their behavior farther from the surface, implying interactions over rather large distances.

To explore these interactions further, we have studied the behavior of large solutes, nominally 1-μm diameter, in the vicinity of hydrophilic surfaces. These large colloidal "solutes" are visually detectable by using ordinary optical microscopy. For hydrophilic surfaces, we examined several common hydrogels because of their anticipated strong interaction with water. We find, unexpectedly, that the solutes are excluded from the vicinity of gel surfaces on a scale of tens of micrometers, and in extreme cases, up to 0.25 millimeters.

## Methods

To explore the behavior of solutes in the vicinity of hydrophilic surfaces, coated latex microspheres were studied in the vicinity of polyvinyl alcohol gels. Two experimental configurations were used (**Fig. 1**). In the first configuration (*A*), a small gel sample was placed between two large glass cover slips and squeezed gently to assure firm contact. Regions peripheral to the gel were filled with a suspension of microspheres, and the chamber was sealed with Parafilm. The assembly was placed on the stage of an inverted microscope (Zeiss Axiovert 35) and viewed in bright field, generally with a 20x objective.

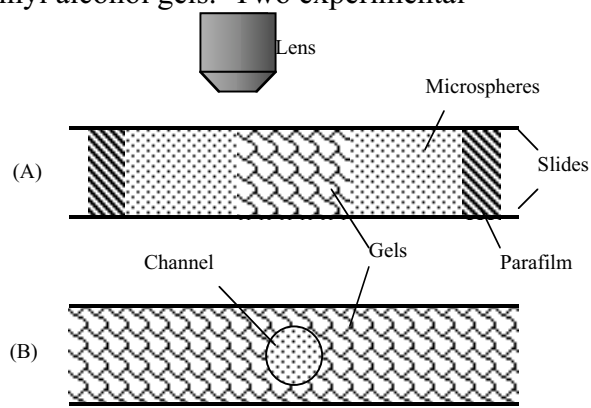

In the second configuration (*B*), the gel was formed around a glass cylinder. Following gelation, the cylinder was withdrawn, leaving a channel 1 mm in diameter. The channel was filled with an aqueous suspension of microspheres and viewed with the optical axis perpendicular to the cylinder axis.

Fig. 1. Methods used to study near-surface effects. *(A)* Gel sample, surrounded by microsphere-containing solution, is sandwiched between two thin glass cover slides, sealed with Parafilm. *(B)* Gel sample contains a solution-filled cylindrical lumen. In both configurations, microspheres are excluded from the region near the gel surface.

Polyvinyl alcohol (PVA) gels were prepared by alternate freezing and thawing of a 3/7 mixture of 10% by wt. PVA solution in water and 10% by wt. PVA solution in dimethyl sulfoxide (DMSO). The mixed solution was injected into a mold to retain the shape either in the form of a rod 0.5 mm in diameter (for configuration *A*), or as a rectangular cube with a 1-mm cylindrical hole (for configuration *B*). Solutions were stored in a freezer (–20˚C) for 23 hrs for physical cross-linking, and then exposed to air at room temperature for 1 hr of annealing. This cycle was repeated four times. Finally the gels were purified by five alternating cycles of immersion in acetone and pure water, and then stored in a large bath of pure water for at least two days. The resulting gels were transparent and had almost the same index of refraction as water.

Polycarboxylate-coated and surfactant-free white aldehyde/amidine-coated microspheres were purchased respectively from Polysciences (Warrington, PA) and Interfacial Dynamics (Tualatin, OR), and kept in the refrigerator until diluted for use in the experiments.

Microscopy was carried out on the stage of an inverted microscope (Zeiss Axiovert 35) and samples were viewed in bright field, generally with a 20x objective. Experiments were carried out at room temperature, and results were recorded on videotape and/or a computer disk.

## Results

In both experimental models, microspheres were distributed nonuniformly (**Fig. 2**). Microspheres were almost completely excluded from the region near the gel surface. Far from the surface, microspheres appeared to be distributed uniformly, and underwent rapid thermal motion. The boundary between exclusion and non-exclusion was typically sharp—on the order of 10% of the width of the exclusion zone. For 2-μm carboxylate microspheres in pure water, the exclusion-zone width was typically 100 μm in the sandwich configuration, and 60 μm in the cylinder configuration.

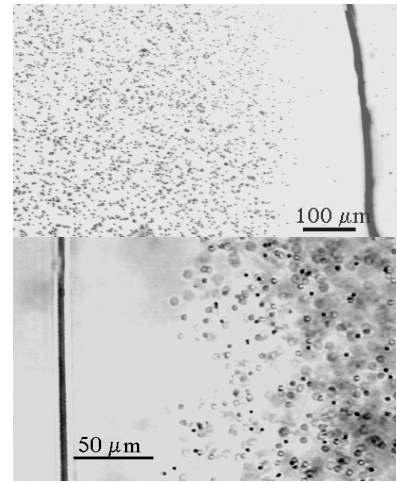

Fig. 2. Results obtained using configuration *A*, top, and configuration *B*, bottom.

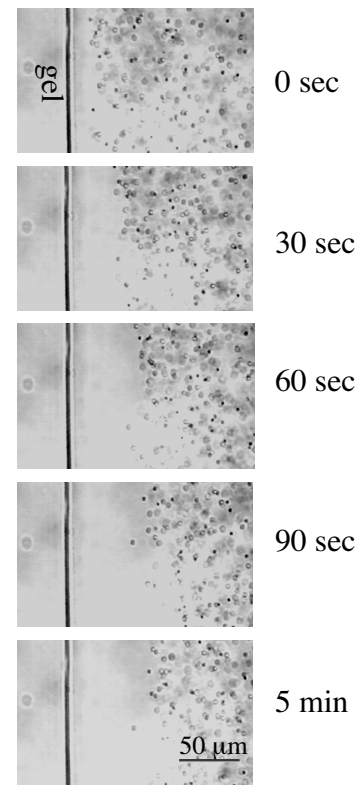

Fig. 3. Development of exclusion zone as function of time in cylindrical channel of PVA gel. Carboxylate-coated microspheres, 2 μm in diameter, were used.

*Potential artifacts*

A number of technical artifacts were considered to account for the observed exclusion. One possible explanation is inadequate diffusion time. If the gel retains a layer of adherent surface water, then microspheres in the newly infiltrated suspension might not have sufficient time to diffuse toward the gel surface during the few minutes between infusion and observation. We found, however, that exclusion zones could persist in samples examined up to a day and sometimes more than a week following infusion (although by that time some of the microspheres had settled to the bottom surface, stuck there, and ceased thermal motion). Furthermore, microspheres situated near the gel surface immediately after infusion migrated away rapidly (**Fig. 3**). Migration velocity in the experiment of Figure 3 was ~1.5 μm per second — fast enough to imply that diffusion was not a limiting factor.

Another possibility is that some polymer strands could project invisibly from the gel proper, perhaps out to 100 μm, creating a zone in which microspheres might be excluded. Although theoretically possible, AFM analysis of various gel surfaces, including the polyacrylamide gel, which also showed exclusion (see below), reveal no evidence of any such projecting polymers (Suzuki et al., 1996). Rather, surfaces are smooth to within 0.2 μm. Furthermore, Figure 3 shows that the zone did not permanently exclude microspheres.

A more conclusive test of the projecting polymer-strand hypothesis was to replace PVA gels that were physically cross-linked with those that were chemically cross-linked, the more stable chemical linkages anticipated to diminish the likelihood that loose strands might emanate from the surface. Thus, physically crosslinked PVA gels were treated in 0.2 wt% glutaraldehyde solution for one day, reacted in the presence of HCl at 30° C for 1 hour, and then purified in the same way as the physically cross-linked PVA gel. Within our resolution, the exclusion zone, measured with 2-μm carboxylate microspheres, was essentially unaffected.

Finally, if such polymer strands were present in sufficiently high density to exclude essentially all 1-μm microspheres, attempts to bend such a bristle-like array would result in macroscopic scale forces. This possibility was tested by placing a deflectable nanolever expanded-tip force probe (Fauver et al., 1998) in a plane parallel to the gel surface, and running the lever in that plane in the direction perpendicular to the lever's long axis. The probe used was capable of detecting forces as low as ~1 pN (Dunaway et al., 2002; Liu et al., 2002). At gel-probe separations ranging from 5 μm to 100 μm, no forces above the 1-pN noise floor were detectable.

Another trivial explanation for the exclusion zone is that the gel shrinks continuously, the outflow of water pushing the microspheres away from the surface and thereby creating an exclusion zone. It is clear from **Figure 3**, however, that the gel boundary does not shift appreciably as the microspheres migrate. More detailed shrinkage analysis was undertaken using samples in configuration *B*. Cylinder diameter was made small enough (148 μm) that both edges could be viewed simultaneously. Over a period of 120 minutes, the dimensional change was typically less than 2 μm and inconsistent in sign. Gel stability was also checked by examining long (10 cm) gels, where length changes of 1% or less could be measured with high accuracy. Again, less than 2% variation was observed over a period of two hours. During the two-minute period of the exclusion transient, then, shrinkage was quantitatively insignificant.

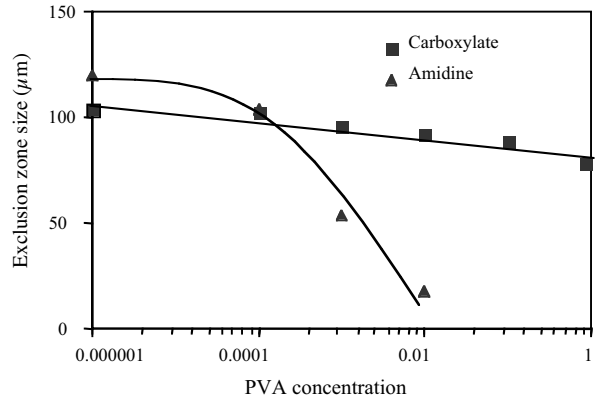

Fig. 4. Effect of PVA molecule concentration on exclusion-zone size using 2-$\mu$m carboxylate and 1.5-$\mu$m amidine microspheres, which were suspended in PVA solutions at pH 10 and pH 2.5, respectively. Curves drawn as guide.

Another possibility is that polymer diffuses out of the gel, creating an invisible polymeric suspension or weak gel near the surface, which excludes large solutes. To test the influence of the presence of polymer molecules in the spheres solution, PVA was added to the microsphere suspension in various concentrations to determine whether exclusion was promoted. We found the opposite. If anything, the presence of polymers tended to reduce the size of the exclusion zone (**Fig. 4**).

As a further test of the polymer-diffusion hypothesis, we explored the behavior of contact-lens gel (polyHEMA, hydroxyethyl methacrylate), which is known to be extremely stable. With 2-μm amidine microspheres, an exclusion zone of ~120 μm was found.

Yet another test of the polymer-diffusion hypothesis was to continuously infuse microsphere suspensions into the cylindrical lumen under pressure (configuration *B*). The measured speeds, reckoned along the cylindrical axis, were up to ~ 100 mm/sec. Any suspended solutes ought to have been swiftly washed out; yet the exclusion zones persisted, virtually unchanged in dimension even at the highest speeds.

(This experiment also rules out the potential for thermal gradients to somehow mediate microsphere exclusion, for the continual high-speed infusion of room-temperature solution should have reduced or eliminated any local gradients.)

Yet another possible explanation is that the exclusion zone arises out of some quirk of the particular gel that was used. The PVA gel was convenient because its transparency permitted visualization of the internal channel in configuration *B*. We checked the generality of the result by substituting several gels beyond the polyHEMA mentioned above. Polyacrylamide gels (configuration *A*) gave qualitatively similar results: for 2-μm carboxylate microspheres, the exclusion-zone width was ~100 μm. Agarose gels also showed similar exclusion — ~60 μm with 1.5-μm amidine microspheres at pH 4.0. We also tried a biological gel: a 200-μm-wide bundle of rabbit-psoas muscle examined in standard physiological buffer. Again, a large exclusion zone was found, ~ 80 μm for 1-

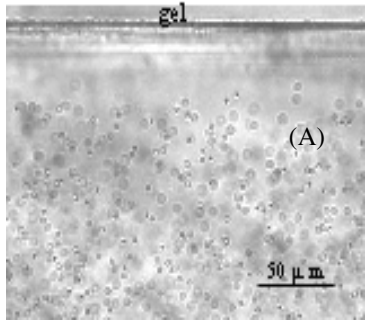

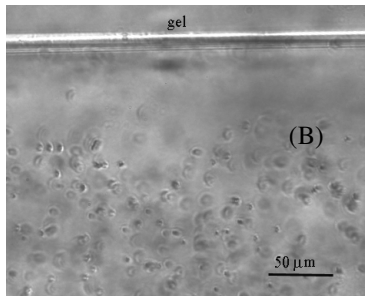

Fig. 5. Comparison of results obtained with microspheres of opposite charge. (A) 2-$\mu m$ carboxylate microspheres, negatively charged. (B) 1.5-$\mu m$ aldehyde/amidine-coated surface, positively charged.

$\mu$m carboxylate microspheres, although in this case the exclusion / non-exclusion boundary was somewhat less sharp than with the artificial gels. Thus, exclusion is not a particular quirk of the PVA gel; it is a general feature associated with various hydrophilic surfaces. On the other hand, not all gels showed exclusion: when polyacrylamide was copolymerized with a vinyl derivative of malachite green, a bulky photoactivatable functional group, no exclusion zone was apparent.

*Characteristics of Exclusion*

A series of experiments was carried out to explore the basic features of the phenomenon. The behaviors of negatively charged carboxylate microspheres and positively charged aldehyde/amidine (surfactant-free white polystyrene latex) microspheres were compared under a variety of experimental conditions. Charge polarity was confirmed by placing the respective microspheres in water, applying a potential difference across the microsphere field, and observing the migration direction.

**Figure 5** shows that exclusion was observed irrespective of whether the microspheres were positively charged or negatively charged. This observation would seem to argue against a simple electrostatic origin of the exclusion. However, we noted that the pH of the water used to dilute the microspheres was sometimes inconsistent because of exposure to air, lending uncertainty to the magnitude of microsphere charge. Hence, the effects of pH were studied systematically with both carboxylate and amidine microspheres.

The effects of pH are shown in **Figure 6**. The PVA gel was stored in distilled water whose pH was measured to be 5.7. Microspheres were added into water adjusted to different pH by addition of HCl or NaOH. For carboxylate microspheres, maximum exclusion was found near the highest pH studied, whereas for amidine microspheres, exclusion was maximum near the lowest pH. Hence, exclusion was most profound essentially when the microspheres were most highly charged.

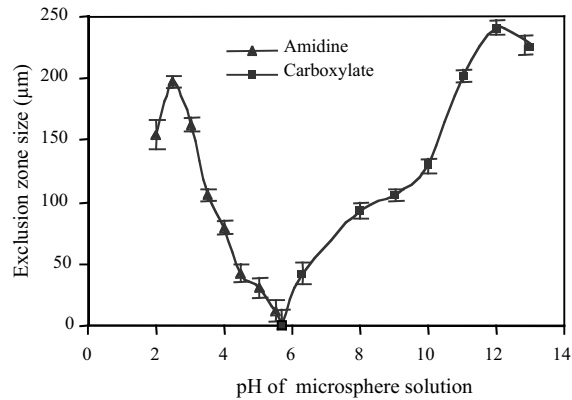

Fig. 6. Effect of pH on size of exclusion zone, where 2-$\mu$m carboxylate microspheres and 1.5-$\mu$m amidine microspheres were suspended in aqueous solution at different pH. PVA gels were stored in pure water at pH 5.7. Note: microspheres of slightly different size were used because the same sizes were not available in amidine and carboxylate. Curves drawn to guide the eye.

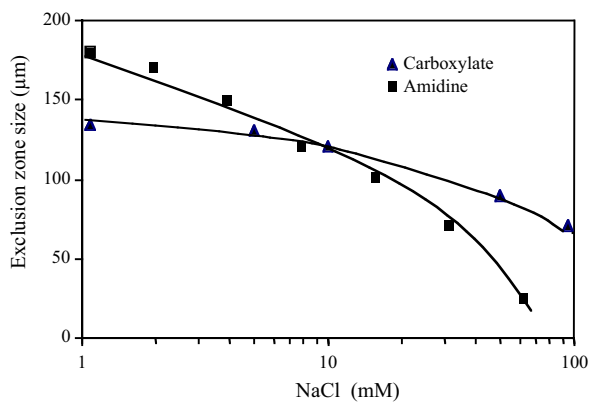

Fig. 7. Effect of salt on exclusion-zone size with 2-$\mu$m carboxylate and 1.5-$\mu$m amidine microspheres suspended in aqueous solutions at pH 2.5 and pH 10, respectively.

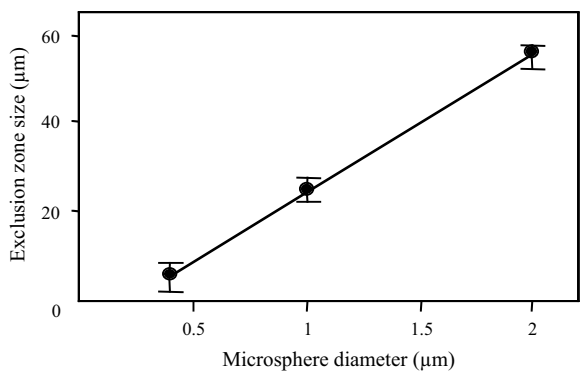

Fig. 8. Effect of solute size on size of the exclusion zone. Carboxylate microspheres.

Effects of salt concentration are shown in **Fig. 7.** The presence of NaCl decreased the size of the exclusion zone. In the case of amidine microspheres, the decrease was moderate, whereas in the case of carboxylate microspheres, it was considerably more shallow, and even up to 100 mM NaCl, there was only a modest reduction in exclusion-zone size.

Solute diameter also played a role in the size of the exclusion zone. **Figure 8** shows a representative result. Here, the relationship between exclusion-zone size and microsphere diameter was seemingly linear within the range studied. The range of usable microsphere size was bounded on the low end by ready visibility, and on the high end by the propensity for microspheres to settle because of gravity.

The experiments were repeated with microspheres of different size suspended simultaneously in the same solution (**Fig. 9**). After equilibrium was reached, the size of the exclusion-zone for each microsphere type was measured from the video image. The dependence of exclusion-zone size on microsphere diameter remained evident, although it did not appear to be linear at all pH values.

The effect of solute concentration was also studied. **Fig. 10** shows that within the concentration range practical to study, exclusion-zone size was virtually independent of microsphere concentration.

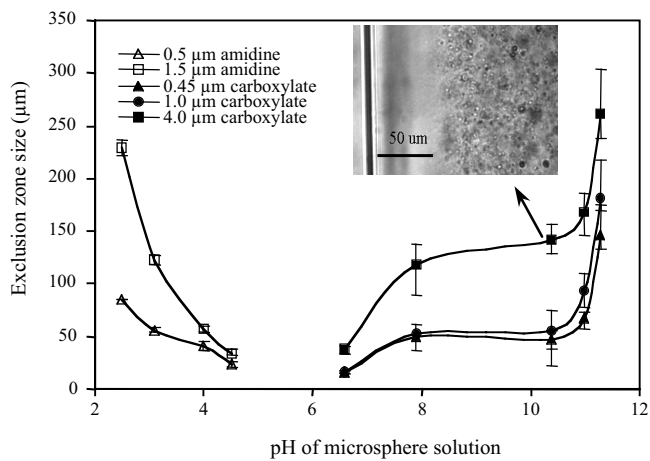

Fig. 9. Effect of microsphere size on exclusion-zone size, measured in microsphere mixtures.

*Exclusion dynamics*

Given the videos showing the development of exclusion, it was possible to measure the time course of microsphere translation away from the surface. This was carried out not only for microspheres initially nearest the gel surface, but also for those initially farther from the surface, including those initially lying beyond the locus of the ultimate exclusion-zone boundary.

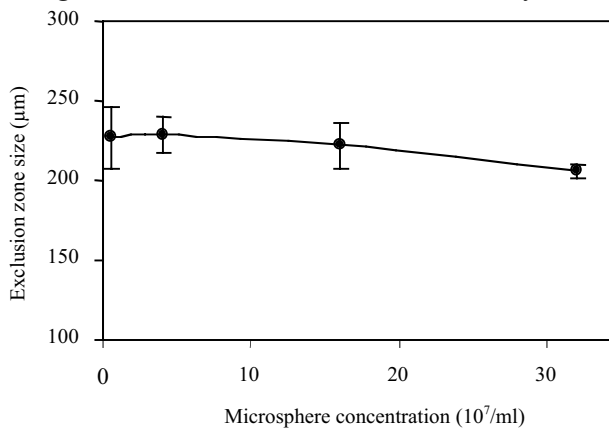

Fig. 10. Effect of concentration of 2-μm amidine microspheres at pH 2.5 on exclusion-zone size.

All microspheres underwent translation, whether initially near to or farther from the gel surface (**Fig. 11**). In the case of amidine microspheres (left), velocity decreased progressively with time, until the microspheres stopped, although it could remain fairly constant for periods of time. Dependence on initial distance from the gel surface was relatively small (upper vs. lower curves).

For carboxylate microspheres (Fig. 11, right), velocity was even more uniformly constant, both temporally and spatially. Those microspheres initially situated beyond the position of the ultimate exclusion boundary translated away from the gel surface at the same velocity as those close to the surface.

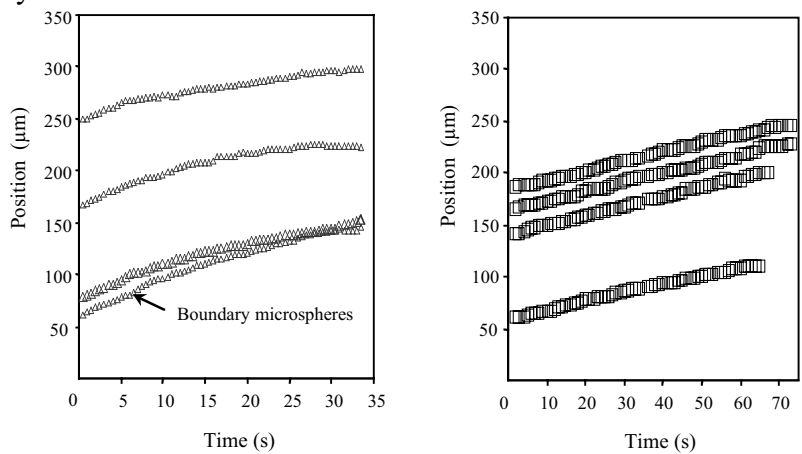

Fig. 11. Time course of exclusion of 1.5-μm amidine microspheres in solution at pH 2.5, (left) and 2-μm carboxylate microspheres in solution at pH 11 (right), after exposure to the PVA gel surface. Different traces show behavior of microspheres at different initial separation from gel.

## Discussion

The finding of a solute-exclusion zone on the order of many tens and up to hundreds of microns is unanticipated. No previous report of surface effects extending for such large distances could be found.

Because the observation itself is quite simple, the first explanation that inevitably comes to mind is that there must be a trivial basis. All artifacts that could be envisioned were checked, some with multiple experiments. We considered: inadequate time for microspheres to diffuse to the gel surface; tethered polymeric strands extending from the gel and creating an exclusion zone; polymers diffusing from the gel and mediating exclusion; thermal gradients creating exclusion; gel shrinkage causing solvent leakage that pushed microspheres from the gel surface; some unknown quirk of the particular gel or solute; and, some consequence of the particular experimental configuration (two different ones were used). All tests proved negative for artifact. The results at hand thus imply that solutes are genuinely excluded from the vicinity of many hydrophilic gel and gel-like (biological) surfaces, for distances on the order of 100 μm for the micron-size solutes studied here.

*Mechanism*

The exclusionary force could originate from at least three potential sites: electrostatic charges, chemical gradients, and water structuring.

In the electrostatic hypothesis, gel and microspheres repel one another and thereby create the exclusion zone. Some electrostatic influence is certainly implied by the pH data (Fig. 9): The exclusion zone was maximum at pH values at which microsphere charge would be expected to have been close to maximum.

On the other hand, other results appeared less consistent with the electrostatic hypothesis. One was the effect of salt (Fig. 7). When NaCl was added to the bath, the size of the exclusion zone decreased, as expected; however, the magnitude of the decrease was very much less than expected. According to standard DLVO theory, the electrostatic potential between surface and solution is thought to drop off as $e^{-X/D}$, where X is distance from the surface and D is the Debye length of the solution (*cf.* Israelachvili, 1992). In ultrapure water, the Debye length may be as large as 1 μm. In a 100 mM salt solution, the Debye length is ~ 1 nm, and at 150 mM salt, a $10^{56}$-fold potential drop is expected at 100 nm. Hence, the implication of appreciable electrostatic force at 100 μm in high salt is far out of accord with standard theoretical predictions. A similar conclusion applies in the case of the biological sample, which showed appreciable exclusion in standard 150 mM buffer.

Another relevant consideration vis-à-vis the electrostatic hypothesis has to do with exclusion dynamics. In the electrostatic hypothesis, microsphere and gel surface are presumed to have the same charge polarity; repulsive forces push the two entities apart until the force is sufficiently small that it can no longer overcome viscous forces, at which point the microsphere ceases to translate. This argument implies that any microsphere situated beyond the final exclusion boundary will experience no net force; yet, microspheres hundreds of microns beyond the boundary did translate, often at the same velocity as those very near the gel (Fig 11, right). This observation also seems difficult to reconcile with a purely electrostatic mechanism.

A second hypothesis is that some pH gradient is set up between gel surface and boundary layer. This hypothesis stems from the observation that a pH difference between gel water and microsphere water seemed a necessary condition for exclusion (Fig. 6). In the absence of the pH gradient, exclusion would vanish and the microspheres would sometimes stick to the gel surface. These pH results imply that an electric field could be built up by a pH difference between the water inside and outside the gel – similar to the liquid-junction potentials that can be observed at liquid-liquid boundaries of electrolyte solutions as a result of the difference of diffusion efficiencies of constituent ions.

Several observations are difficult to explain in these terms. First, liquid-junction potentials extend only over very small distances. Second, any such gradient would be expected to diminish with time; yet, the exclusion zone persisted easily for hours. Third, local translation velocity would be expected to be proportional to the local pH gradient, but Fig. 11 shows that velocity could be independent of distance from the gel surface.

A third hypothesis is that the exclusion is caused by layers of water molecules growing in an organized manner from the gel surface. Layers of tightly bound water are known to exist around hydrophilic polymers, either charged or polar (Vogler, 1998). Although their extent is unclear, the number of layers is thought to lie in the single digits, but the experiments of Pashley and Kitchener (1979) and others (Fisher et al., 1981; Xu and Yeung, 1998) leave open the possibility of more substantial layering. Layers could build one upon another, beginning at the gel surface and extending outward, excluding solutes as the number of layers grows.

On the other hand, no reports we know of suggest any more than several hundred layers in the extreme. Hence, the current result implying up to $10^6$ solvent layers would be unprecedented. In liquid crystals, however, molecular alignment of small solvent molecules is recognized to occur over macroscopic distances (Arabia et al., 1991; Marrink et al., 1996). Also, in clouds, it is well known that water is clustered around condensation nuclei, forming aerosol droplets on the order of microns. Hence, this line of interpretation may have some precedent.

If the exclusionary force does arise from structured water, one expectation is that in the steady state, larger solutes should be more profoundly excluded than smaller ones, and this has been confirmed for various sugars and other low molecular weight solutes (Ling et al., 1993). This feature was confirmed here for larger solutes (Figs. 8, 9).

Another relevant point is translation dynamics. Microsphere-translation velocity was relatively constant as a function of time over much of the course of its excursion. If the laying of one water stratum depends on the presence of exposed charge of a previous stratum, relatively constant velocity is anticipated. Also, the observed sharpness of the steady-state exclusion boundary would seem consistent with this hypothesis.

The effect of pH is also relevant. Exclusion was greatest when microsphere charge was greatest (Fig. 6). Higher surface charge is known to be associated with larger extents of water structuring (Vogler, 1998). Thus, within the framework of the water-structure hypothesis, exclusion would be attributed to the combined structuring capacity of gel surface and microsphere surfaces. Both zones of structure would repel one another. Water structuring between gel surfaces might explain why the microspheres moved almost as a unit, with microsphere velocity near the gel surface similar to that far from the gel surface (Fig. 11).

In sum, the mechanism of exclusion is not yet clear, and considerably more work in spectroscopic and other approaches will be required before it can be settled. Electrostatic and pH-gradient hypotheses seem excluded by several observations. And, while the water structure hypothesis seems *a priori* most consistent with at least some of the evidence, the implication of extensive water structuring, orders of magnitude beyond what is generally thought, would be an astonishing conclusion. Perhaps the mechanism lies outside any of these hypotheses.

The presence of such long-range forces between solutes could have profound significance for molecular interactions in both natural and artificial systems (Pollack, 2001). Interactions are anticipated to occur on the nanometer scale, but the results imply that interactions are possible on a scale of tens, or even hundreds, of micrometers.

### Acknowledgments

We thank Drs. Xiumei Liu, Yudong Hao, and John Krieger for help with preliminary experiments, Dr. Hitoshi Suda for performing the nanolever scans, Dr. Angela Carden for consistent advice along the way, and Prof. Toshio Hirai, who first alerted us to possible solute exclusion in his studies of flow profiles. For constructive comments on earlier versions of the manuscript, we thank Philip Ball, Mark Banaszak Holl, Frank Borg, James Clegg, Evan Evans, Jacob Israelachvili, Wolfgang Linke, Adrian Parsegian, Mickey Schurr, Erwin Vogler, John Watterson and Philippa Wiggins.